# Flexible Ansatz for *N*-body Perturbation Theory

Ramón Alain Miranda-Quintana,[1*] Taewon D. Kim,[1,2] Rugwed A. Lokhande,[1] M. Richer,[2] Gabriela Sánchez-Díaz,[2] Pratiksha B. Gaikwad,[1] Paul W. Ayers,[2*]

1. Department of Chemistry and Quantum Theory Project, University of Florida, Gainesville, FL 32603, USA
2. Department of Chemistry and Chemical Biology, McMaster University, Hamilton, Ontario, L8S 4M1, Canada

Emails: ayers@mcmaster.ca, quintana@chem.ufl.edu

**Abstract**

We propose a new Perturbation Theory framework that can be used to help with the projective solution of the Schrödinger equation for arbitrary wavefunctions. This Flexible Ansatz for *N*-body Perturbation Theory (FANPT) is based on our previously proposed Flexible Ansatz for *N*-body Configuration Interaction (FANCI). We derive recursive FANPT expressions including arbitrary orders in the perturbation hierarchy. We show that the FANPT equations are well-behaved across a wide range of conditions, including static correlation-dominated configurations and highly non-linear wavefunctions.

## 1. INTRODUCTION

In the first paper of this series, we presented a generalization of the Configuration Interaction (CI) method for solving the Schrödinger equation. The key difference to the usual CI is that we allow the Slater determinant coefficients of the wavefunction to be functions of certain parameters, $\vec{P} = [p_1, p_2, ...]$:

$$\left| \Psi(\vec{P}) \right\rangle = \sum_{\mathbf{m} \in S} f_{\mathbf{m}}(\vec{P}) | \mathbf{m} \rangle \qquad (1)$$

where $|\mathbf{m}\rangle$ is a Slater determinant in the (possibly restricted) space of Slater determinants $S$, as specified by the occupation numbers, $\mathbf{m}$, of the spin-orbitals (single-particle states), which we assume to be orthogonal and normalized. According to this, the function $f_{\mathbf{m}}(\vec{P})$ corresponds to the overlap between the wavefunction ansatz and an electronic configuration:[1] $f_\mathbf{m}(\vec{P}) = \langle \mathbf{m} | \Psi(\vec{P}) \rangle$. We refer to Eq. (1) as the Flexible Ansatz for *N*-body Configuration Interaction (FANCI). As demonstrated elsewhere, this formulation includes not only conventional CI[2-12] and Coupled Cluster[13-19] (CC) approaches, but also Matrix-Product States,[20-27] Tensor Network States,[28-32] and a wide variety of wavefunctions based on composite bosons and composite fermions, including many flavors of geminals.[33-53]

For certain ansatze, it is feasible to find the parameters and energy, *E*, corresponding to the wavefunction in Eq. (1) using a variational approach. In some cases, it is even convenient to generalize Eq. (1) so that the functions used in the expansion are not Slater determinants, but more complicated objects like Configuration State Functions[3,54-56] (CSFs) or the eigenfunctions of model Hamiltonians.[57-61] However, more generally, we find the wavefunction parameters and *E* by using the projected Schrödinger equation:

$$\sum_{\mathbf{m} \in S} f_\mathbf{m}(\vec{P}) \langle \mathbf{n} | \hat{H} | \mathbf{m} \rangle = E f_\mathbf{n}(\vec{P}) \qquad |\mathbf{m}\rangle \in P \qquad (2)$$

over a suitably defined projection space, *P*.

To solve non-linear equations like Eq. (2), it is very helpful to have a reasonable guess for the solution. For example, in traditional CC calculations, an initial guess for the cluster amplitudes is obtained from Møller-Plesset perturbation theory.[62-70] In this paper, we will present a perturbative approach[14,71-78] to the projective Schrödinger equation for completely general FANCI wavefunctions. We call the resulting method a Flexible Ansatz for *N*-body Perturbation Theory, FANPT. In particular, we will consider the case in which we have the same number of equations and unknowns in Eq. (2), while assuming (for simplicity), that the wavefunction, wavefunction parameters, and Slater determinant coefficients are all real. The extension of our results to complex-valued functions is perfectly straightforward, while the analysis of over-determined systems of equations will be presented in a forthcoming contribution.

## 2. Flexible Ansatz for *N*-body Perturbation Theory (FANPT)
### A. First-Order FANPT

As usual in perturbation theory (PT), we rewrite the Hamiltonian as a simple-to-solve 0$^\text{th}$-order term, *F*, plus a perturbation, *V*. Then, we can go from the ideal to the real Hamiltonian by changing a perturbation parameter, $\lambda$, from 0 to 1:

$$\hat{H} = \hat{F} + \lambda \hat{V} \tag{3}$$

The 0$^{th}$-order reference operator is usually a one-body Hamiltonian defined by a (possibly generalized) Fock operator. The key assumption is that the projected Schrödinger equation is easy to solve in this case, giving the energy and wavefunction parameters in that limit, $E_0$ and $\vec{P}_0$, respectively. Using Eq. (3), we can turn Eq. (2) into:

$$\sum_{\mathbf{m} \in S} f_{\mathbf{m}}(\vec{P}_\lambda) \langle \mathbf{n} | \hat{F} + \lambda \hat{V} | \mathbf{m} \rangle = E_\lambda f_{\mathbf{n}}(\vec{P}_\lambda) \tag{4}$$

For convenience, we will rewrite this system of non-linear equations in standard form:

$$G_{\mathbf{n}}^{(0)}(\lambda, E_\lambda, \vec{P}_\lambda) = \sum_{\mathbf{m} \in S} f_{\mathbf{m}}(\vec{P}_\lambda) \langle \mathbf{n} | \hat{F} + \lambda \hat{V} | \mathbf{m} \rangle - E_\lambda f_{\mathbf{n}}(\vec{P}_\lambda) = 0 \tag{5}$$

Then, the linear differential of this system of non-linear equations can be written as:

$$\left( \frac{\partial G_{\mathbf{n}}^{(0)}(\lambda, E_\lambda, \vec{P}_\lambda)}{\partial \lambda} \right)_{E_\lambda, \vec{P}_\lambda} d\lambda + \left( \frac{\partial G_{\mathbf{n}}^{(0)}(\lambda, E_\lambda, \vec{P}_\lambda)}{\partial E_\lambda} \right)_{\lambda, \vec{P}_\lambda} dE_\lambda + \sum_k \left( \frac{\partial G_{\mathbf{n}}^{(0)}(\lambda, E_\lambda, \vec{P}_\lambda)}{\partial p_{k;\lambda}} \right)_{E_\lambda, \lambda} dp_{k;\lambda} = 0$$

(6)

which can be re-arranged as a system of linear equations to solve for the derivative of the energy and the wavefunction parameters with respect to $\lambda$:

$$\left( \frac{\partial G_{\mathbf{n}}^{(0)}(\lambda, E_\lambda, \vec{P}_\lambda)}{\partial E_\lambda} \right)_{\lambda, \vec{P}_\lambda} \frac{dE_\lambda}{d\lambda} + \sum_k \left( \frac{\partial G_{\mathbf{n}}^{(0)}(\lambda, E_\lambda, \vec{P}_\lambda)}{\partial p_{k;\lambda}} \right)_{\substack{E_\lambda, \lambda \\ \vec{P}_\lambda \neq p_{k;\lambda}}} \frac{dp_{k;\lambda}}{d\lambda} = -\left( \frac{\partial G_{\mathbf{n}}^{(0)}(\lambda, E_\lambda, \vec{P}_\lambda)}{\partial \lambda} \right)_{E_\lambda, \vec{P}_\lambda}$$

(7)

The coefficients in this system can be easily calculated. For example, at $\lambda = 0$:

$$\left( \frac{\partial G_{\mathbf{n}}^{(0)}(\lambda, E_\lambda, \vec{P}_\lambda)}{\partial E_\lambda} \right)_{\lambda=0, \vec{P}_\lambda} = -f_{\mathbf{n}}(\vec{P}_{\lambda=0}) \tag{8}$$

$$\left( \frac{\partial G_{\mathbf{n}}^{(0)}(\lambda, E_\lambda, \vec{P}_\lambda)}{\partial p_{k;\lambda}} \right)_{\substack{E_\lambda, \lambda=0 \\ \vec{P}_\lambda \neq p_{k;\lambda}}} = \sum_{\mathbf{m} \in S} \langle \mathbf{n} | \hat{F} | \mathbf{m} \rangle \frac{\partial f_{\mathbf{m}}(\vec{P}_\lambda)}{\partial p_{k;\lambda}} \bigg|_{\lambda=0} - E_0 \frac{\partial f_{\mathbf{n}}(\vec{P}_\lambda)}{\partial p_{k;\lambda}} \bigg|_{\lambda=0} \tag{9}$$

$$\left( \frac{\partial G_{\mathbf{n}}^{(0)}(\lambda, E_\lambda, \vec{P}_\lambda)}{\partial \lambda} \right)_{E_\lambda, \vec{P}_\lambda} = \sum_{\mathbf{m} \in S} \langle \mathbf{n} | \hat{V} | \mathbf{m} \rangle f_{\mathbf{m}}(\vec{P}_0) \tag{10}$$

These general expressions can be simplified greatly if we consider the usual case where $\hat{F} = \hat{H}_0$ is a one-body operator, the 0$^{th}$-order wavefunction is a Slater determinant, and the 0$^{th}$-order energy is the sum of occupied orbital energies. Using this result:

$$\hat{F}|\mathbf{n}_0\rangle = E_0 |\mathbf{n}_0\rangle \tag{11}$$

and assuming that the Slater determinants are eigenfunctions of $\hat{F}$:

$$\hat{F}|\mathbf{m}\rangle = E_\mathbf{m}|\mathbf{m}\rangle \tag{12}$$

then Eqs. (8)-(10) transform into:

$$\left(\frac{\partial G_\mathbf{n}^{(0)}(\lambda, E_\lambda, \vec{P}_\lambda)}{\partial E_\lambda}\right)_{\lambda=0, \vec{P}_\lambda} = -\delta_{\mathbf{nn}_0} \tag{13}$$

$$\left(\frac{\partial G_\mathbf{n}^{(0)}(\lambda, E_\lambda, \vec{P}_\lambda)}{\partial p_{k;\lambda}}\right)_{\substack{E_\lambda, \lambda=0 \\ \vec{P}_\lambda \neq p_{k;\lambda}}} = (E_\mathbf{n} - E_0) \left.\frac{\partial f_\mathbf{n}(\vec{P}_\lambda)}{\partial p_{k;\lambda}}\right|_{\lambda=0} \tag{14}$$

$$\left(\frac{\partial G_\mathbf{n}^{(0)}(\lambda, E_\lambda, \vec{P}_\lambda)}{\partial \lambda}\right)_{E_\lambda, \vec{P}_\lambda} = \langle \mathbf{n}|\hat{V}|\mathbf{n}_0\rangle \tag{15}$$

The linear equation can then be rewritten as:

$$-\delta_{\mathbf{nn}_0} \frac{dE_\lambda}{d\lambda} + \sum_k (E_\mathbf{n} - E_0) \left.\frac{\partial f_\mathbf{n}(\vec{P}_\lambda)}{\partial p_{k;\lambda}}\right|_{\lambda=0} \frac{dp_{k;\lambda}}{d\lambda} = -\langle \mathbf{n}|\hat{V}|\mathbf{n}_0\rangle \tag{16}$$

which simplifies further to:

$$\left.\frac{dE_\lambda}{d\lambda}\right|_{\lambda=0} = \langle \mathbf{n}_0|\hat{V}|\mathbf{n}_0\rangle \tag{17}$$

$$\sum_k \left.\frac{\partial f_\mathbf{n}(\vec{P}_\lambda)}{\partial p_{k;\lambda}}\right|_{\lambda=0} \left.\frac{dp_{k;\lambda}}{d\lambda}\right|_{\lambda=0} = \frac{\langle \mathbf{n}|\hat{V}|\mathbf{n}_0\rangle}{(E_0 - E_\mathbf{n})} \quad \mathbf{n} \neq \mathbf{n}_0 \tag{18}$$

So far, we have assumed that the normalization of the wavefunction is specified by fixing one of the wavefunction parameters to have a set value. If this is not the case, an additional equation corresponding to a desired relationship between the wavefunction parameters or intermediate normalization,

$$f_{\mathbf{n}_0}(\vec{P}_\lambda) = 1 \tag{19}$$

should be added to the system of equations defined in Eq. (5). For the case of intermediate normalization, this means adding the following equation:

$$\sum_k \frac{\partial f_{\mathbf{n}_0}(\vec{P}_\lambda)}{\partial p_{k;\lambda}} \frac{dp_{k;\lambda}}{d\lambda} = 0 \qquad (20)$$

**B. Second-Order FANPT**

The second order of FANPT is obtained by recognizing that the 1$^{st}$-order PT equations are, in fact, another system of equations to solve. Hence, we can write Eq. (7) as:

$$G_{\mathbf{n}}^{(1)} \equiv \left(\frac{\partial G_{\mathbf{n}}^{(0)}(\lambda, E_\lambda, \vec{P}_\lambda)}{\partial E_\lambda}\right)_{\lambda, \vec{P}_\lambda} \frac{dE_\lambda}{d\lambda} + \sum_k \left(\frac{\partial G_{\mathbf{n}}^{(0)}(\lambda, E_\lambda, \vec{P}_\lambda)}{\partial p_{k;\lambda}}\right)_{\substack{E_\lambda, \lambda \\ \vec{P}_\lambda \neq p_{k;\lambda}}} \frac{dp_{k;\lambda}}{d\lambda} + \left(\frac{\partial G_{\mathbf{n}}^{(0)}(\lambda, E_\lambda, \vec{P}_\lambda)}{\partial \lambda}\right)_{E_\lambda, \vec{P}_\lambda} = 0$$

(21)

Then, we can repeat the argument from the previous session, obtaining:

$$\left(\frac{\partial G_{\mathbf{n}}^{(0)}(\lambda, E_\lambda, \vec{P}_\lambda)}{\partial E_\lambda}\right)_{\lambda, \vec{P}_\lambda} \left.\frac{d^2 E_\lambda}{d\lambda^2}\right|_{\lambda=0} +$$

$$\sum_k \left(\frac{\partial G_{\mathbf{n}}^{(0)}(\lambda, E_\lambda, \vec{P}_\lambda)}{\partial p_{k;\lambda}}\right)_{\substack{E_\lambda, \lambda \\ \vec{P}_\lambda \neq p_{k;\lambda}}} \left.\frac{d^2 p_{k;\lambda}}{d\lambda^2}\right|_{\lambda=0} = -\left\{\begin{array}{l} \sum_{k,l}\left(\dfrac{\partial^2 G_{\mathbf{n}}^{(0)}(\lambda, E_\lambda, \vec{P}_\lambda)}{\partial p_{k;\lambda}\partial p_{l;\lambda}}\right)_{\substack{E_\lambda, \lambda \\ \vec{P}_\lambda \neq p_{k;\lambda}, p_{l;\lambda}}} \left.\dfrac{dp_{k;\lambda}}{d\lambda}\right|_{\lambda=0} \left.\dfrac{dp_{l;\lambda}}{d\lambda}\right|_{\lambda=0} \\ +2\sum_k \left(\dfrac{\partial^2 G_{\mathbf{n}}^{(0)}(\lambda, E_\lambda, \vec{P}_\lambda)}{\partial E_\lambda \partial p_{k;\lambda}}\right)_{\lambda, \vec{P}_\lambda \neq p_{k;\lambda}} \left.\dfrac{dp_{k;\lambda}}{d\lambda}\right|_{\lambda=0} \left.\dfrac{dE}{d\lambda}\right|_{\lambda=0} \\ +2\sum_k \left(\dfrac{\partial^2 G_{\mathbf{n}}^{(0)}(\lambda, E_\lambda, \vec{P}_\lambda)}{\partial \lambda \partial p_{k;\lambda}}\right)_{E_\lambda, \vec{P}_\lambda \neq p_{k;\lambda}} \left.\dfrac{dp_{k;\lambda}}{d\lambda}\right|_{\lambda=0} \end{array}\right\}$$

(22)

Notice that the linear coefficients of the unknown 2$^{nd}$-order terms are unchanged from the 1$^{st}$-order Eq. (7). The term on the right-hand-side has changed, however, and is now significantly more complicated. In the special case where the 0$^{th}$-order wavefunction is a Slater determinant, this equation reduces to:

$$\left.\frac{d^2 E_\lambda}{d\lambda^2}\right|_{\lambda=0} = 2\left\{\begin{array}{l} \sum_k \left.\dfrac{\partial f_{\mathbf{n}_0}(\vec{P}_\lambda)}{\partial p_{k;\lambda}}\right|_{\lambda=0} \left.\dfrac{dp_{k;\lambda}}{d\lambda}\right|_{\lambda=0} \left.\dfrac{dE}{d\lambda}\right|_{\lambda=0} \\ +\sum_{\mathbf{m}}\sum_k \langle \mathbf{n}_0|\hat{V}|\mathbf{m}\rangle \left.\dfrac{\partial f_{\mathbf{m}}(\vec{P}_\lambda)}{\partial p_{k;\lambda}}\right|_{\lambda=0} \left.\dfrac{dp_{k;\lambda}}{d\lambda}\right|_{\lambda=0} \end{array}\right\} \qquad (23)$$

$$\sum_k \left.\frac{\partial f_\mathbf{n}(\vec{P}_\lambda)}{\partial p_{k;\lambda}}\right|_{\lambda=0} \left.\frac{d^2 p_{k;\lambda}}{d\lambda^2}\right|_{\lambda=0} = \left\{\begin{array}{l} -\sum_{k,l} \left.\frac{\partial^2 f_\mathbf{n}(\vec{P}_\lambda)}{\partial p_{k;\lambda} \partial p_{l;\lambda}}\right|_{\lambda=0} \left.\frac{dp_{k;\lambda}}{d\lambda}\right|_{\lambda=0} \left.\frac{dp_{l;\lambda}}{d\lambda}\right|_{\lambda=0} \left.\frac{dE}{d\lambda}\right|_{\lambda=0} \\ +\frac{2}{E_\mathbf{n}-E_0} \sum_k \left.\frac{\partial f_\mathbf{n}(\vec{P}_\lambda)}{\partial p_{k;\lambda}}\right|_{\lambda=0} \left.\frac{dp_{k;\lambda}}{d\lambda}\right|_{\lambda=0} \left.\frac{dE}{d\lambda}\right|_{\lambda=0} \\ -\frac{2}{E_\mathbf{n}-E_0} \sum_\mathbf{m} \sum_k \langle\mathbf{n}|\hat{V}|\mathbf{m}\rangle \left.\frac{\partial f_\mathbf{m}(\vec{P}_\lambda)}{\partial p_{k;\lambda}}\right|_{\lambda=0} \left.\frac{dp_{k;\lambda}}{d\lambda}\right|_{\lambda=0} \end{array}\right\} \quad \mathbf{n} \neq \mathbf{n}_0$$

(24)

These equations are derived in the Supplementary Information.

**C. Approximate Higher-Order FANPT**

Higher-order corrections can be derived, recursively, by the same strategy. The primary term that affects the cost is the need to evaluate higher-order derivatives of $f_\mathbf{m}(\vec{P}_\lambda)$ with respect to its parameters. If we *assume* that the 2$^\text{nd}$- and higher-order derivatives $\frac{\partial^n f_\mathbf{m}(\vec{P}_\lambda)}{\partial p_{k_1;\lambda} \partial p_{k_2;\lambda}\cdots}$ are negligible, and that the 0$^\text{th}$-order Hamiltonian is a one-body operator that satisfies Eqs. (11) and (12), then the higher-order perturbations are simply determined by the recursions:

$$\left.\frac{d^n E_\lambda}{d\lambda^n}\right|_{\lambda=0} = \sum_\mathbf{m}\sum_k \langle\mathbf{n}_0|\hat{V}|\mathbf{m}\rangle \left.\frac{\partial f_\mathbf{m}(\vec{P}_\lambda)}{\partial p_{k;\lambda}}\right|_{\lambda=0} \left(n \left.\frac{d^{n-1} p_{k;\lambda}}{d\lambda^{n-1}}\right|_{\lambda=0}\right) \\ -\sum_k \left.\frac{\partial f_{\mathbf{n}_0}(\vec{P}_\lambda)}{\partial p_{k;\lambda}}\right|_{\lambda=0} \left\{\sum_{m=1}^{n-1}\binom{n}{m}\left.\frac{d^m E}{d\lambda^m}\right|_{\lambda=0} \left.\frac{d^{n-m} p_{k;\lambda}}{d\lambda^{n-m}}\right|_{\lambda=0}\right\}$$

(25)

$$\sum_k \left.\frac{\partial f_\mathbf{n}(\vec{P}_\lambda)}{\partial p_{k;\lambda}}\right|_{\lambda=0} \left.\frac{d^n p_{k;\lambda}}{d\lambda^n}\right|_{\lambda=0} = -\frac{1}{E_\mathbf{n}-E_0}\sum_\mathbf{m}\sum_k \langle\mathbf{n}|\hat{V}|\mathbf{m}\rangle \left.\frac{\partial f_\mathbf{m}(\vec{P}_\lambda)}{\partial p_{k;\lambda}}\right|_{\lambda=0} \left(n \left.\frac{d^{n-1} p_{k;\lambda}}{d\lambda^{n-1}}\right|_{\lambda=0}\right) \\ \frac{1}{E_\mathbf{n}-E_0}\sum_k \left.\frac{\partial f_\mathbf{n}(\vec{P}_\lambda)}{\partial p_{k;\lambda}}\right|_{\lambda=0} \left\{\sum_{m=1}^{n-1}\binom{n}{m}\left.\frac{d^m E}{d\lambda^m}\right|_{\lambda=0} \left.\frac{d^{n-m} p_{k;\lambda}}{d\lambda^{n-m}}\right|_{\lambda=0}\right\} \quad \mathbf{n}\neq\mathbf{n}_0$$

(26)

Because the linear equations for the perturbed wavefunctions parameters always have the same matrix, the cost of perturbation theory does not increase with order in this approximation. These equations are derived in the Supplementary Information section. Notice that if the matrix representing the derivatives of the coefficients of the Slater determinants in the projection space with respect to the wavefunction parameters $\frac{\partial f_{\mathbf{n}}(\vec{P}_\lambda)}{\partial p_{k;\lambda}}$ is invertible, then one can trivially obtain explicit equations for the higher-order response functions $\frac{d^n p_{k;\lambda}}{d\lambda^n}$, subject to the approximation:

$$n \geq 2 \Rightarrow \frac{\partial^n f_{\mathbf{m}}(\vec{P}_\lambda)}{\partial p_{k_1;\lambda} \partial p_{k_2;\lambda} \ldots} = 0 \qquad (27)$$

This amounts to assuming that the parameterization is linear; i.e., one neglects terms that would not appear in normal configuration interaction calculation, where the wavefunction ansätz is a linear combination of determinants.

One could include, without prohibitive extra computation, the diagonal terms in the derivative tensor, $\partial^n f_{\mathbf{m}}(\vec{P}_\lambda)/\partial p_{k_1;\lambda}^n$. However, these terms, and all the terms other than the "fully off-diagonal" (with no repeated parameters) derivatives are zero in most wavefunction ansatze we have considered, including those based on excitation-based coupled-cluster methods (because repeated excitations vanish) and geminal ansatze (because parameters contribute at most once in each permutation).

## 3. RESULTS

We tested how FANPT can be used to solve the (projective) FANCI equations under different conditions. The overall procedure is quite simple: First, we need to decide in how many "steps" we are going to use to represent the adiabatic connection link the reference Hamiltonian to the real Hamiltonian (from $\lambda = 0$ to $\lambda = 1$ in Eq. (3)). Also, we need to determine up to which order we will include the PT corrections. Once these factors have been decided, we use the wavefunction parameters and $E$ for a value of $\lambda$ as input to the FANPT equations to estimate the parameters and energy at a new value $\lambda + \Delta\lambda$. The FANPT estimate then serves as initial guess to solve the FANCI equations at $\lambda + \Delta\lambda$. That is, FANPT serves as a "guide" providing a more robust starting point to the solution of the FANCI equations throughout the $\lambda = 0 \to \lambda = 1$ path.

| WFN | Steps/Order | 1 | 2 | 3 | 4 | Reference $E$ |
|---|---|---|---|---|---|---|
| *CISD* | 10 | -7.971770 | -7.972082 | -7.972083 | -7.972084 | -7.972084 |
|  | 100 | -7.972083 | -7.971818 | -7.972083 | -7.972082 |  |

|          | 1000 | -7.971803 | -7.971784 | -7.972083 | -7.971800 |           |
|----------|------|-----------|-----------|-----------|-----------|-----------|
| CCSD     | 10   | -7.972235 | -7.972110 | -7.972098 | -7.972115 | -7.971722 |
|          | 100  | -7.972216 | -7.972098 | -7.972491 | -7.972546 |           |
|          | 1000 | -7.972100 | -7.972164 | -7.972122 | -7.972099 |           |
| AP1roGSD | 10   | -7.964657 | -7.964650 | -7.964637 | -7.964638 | -7.967194 |
|          | 100  | -7.964689 | -7.964699 | -7.964676 | -7.964690 |           |
|          | 1000 | -7.964674 | -7.964672 | -7.964689 | -7.964685 |           |

**Table 1:** 1st, 2nd, 3rd, and 4th order FANPT energies (in atomic units) for CISD, CCSD, and AP1roGSD wavefunctions with 10, 100, and 1000 steps for LiH at equilibrium bond distance. Orders 2, 3, and 4 used the quasi-linear approximation, Eq. (27). All calculations used the STO-6G basis.

Table 1 shows that the FANPT equations are well-behaved over a range of PT steps and orders, and for different types of wavefunctions: CI (CISD),[79-85] CC (CCSD),[86-88] and geminals (AP1roGSD).[89] As Eq. (27) is exact for any (selected) CI method, it is no surprise that the agreement between the FANPT and the reference CISD results is perfect. However, it is reassuring that even with highly non-linear wavefunctions like CCSD and AP1roGSD, a seemingly drastic approximation like Eq. (27) still leads to ~ mHartree agreement with the reference energies. A more demanding test is shown in Fig. 1, where it can be seen that FANPT seamlessly converges to the underlying non-linear geminal wavefunctions (APIG[48,49,90] and APG[91,92]) for different orders of PT and over regions where static correlation effects are dominant.

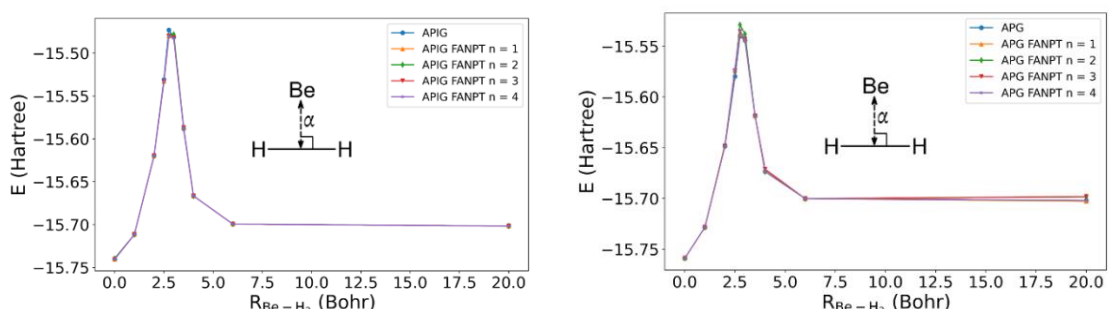

**Figure 1:** Different orders (n = 1-4) of FANPT calculations (100 steps) for the $C_{2v}$ insertion of Be into $H_2$ using APIG (left) and APG (right). Orders 2, 3, and 4 used the quasi-linear approximation, Eq. (27). All calculations used the STO-6G basis. Nuclear configurations were taken from Ref. [93].

## 4. CONCLUSIONS

We proposed FANPT as a way to guide the solution of the projected Schrödinger equations for arbitrary wavefunctions within the FANCI framework. By progressively evolving from a (simple to solve) $0^{th}$ order model Hamiltonian, FANPT effectively provides a robust initial guess for the FANCI equations. Interestingly, when FANPT is applied to standard CI methods, it provides an alternative to traditional diagonalization approaches. While this does not seem more efficient than routine diagonalization algorithms, it might be useful in cases where iterative diagonalization was difficult to converge.

Despite the "quasi-linear" approximation (Eq. (27)), our numerical tests show that FANPT serves as a reliable guide throughout the solution process, facilitating the convergence of FANCI equations under various conditions. Whether applied to conventional CI and CC, or geminal-based wavefunctions, FANPT consistently converged to the underlying reference wavefunction. The agreement between FANPT results and the reference results validates the efficacy of our approach in capturing essential correlation effects and yielding accurate energy predictions over non-trivial electronic configurations. Moving forward, we plan to explore FANPT not only as a (very) valuable complement to FANCI, but as a stand-alone method capable of providing insights into the solution of non-linear equations.

**Acknowledgements:** RAMQ, PBG, RAL, and TDK thank support from ORAU in the form of a Ralph E. Powe award. RAMQ and PBG thank MolSSI for a Software Fellowship. PWA thanks the Canada Research Chairs, NSERC, and the Digital Research Alliance of Canada. PWA acknowledges NSERC, the Canada Research Chairs, and the Digital Research Alliance of Canada.